\documentclass[nofootinbib,twocolumn,preprintnumbers,amsmath,amssymb]{revtex4}
\usepackage[hyperfootnotes=true, linktocpage=true, citecolor=blue, linkcolor=blue, urlcolor=Maroon, filecolor=Maroon]{hyperref}
\usepackage[noconfig]{refstyle}
\usepackage{graphicx}% Include figure files
\usepackage{dcolumn}% Align table columns on decimal point
\usepackage{bm}% bold math

\usepackage{xcolor}
\usepackage{multirow}

\usepackage{enumitem}
\newlist{todolist}{itemize}{2}
\setlist[todolist]{label=$\square$}
\usepackage{pifont}

\usepackage{tikz-cd}

\newcommand{\beq}{\begin{equation}}
\newcommand{\eeq}{\end{equation}}
\newcommand{\bea}{\begin{eqnarray}}
\newcommand{\eea}{\end{eqnarray}}

\newtheorem{theorem}{Theorem}%[section]

\newtheorem{claim}[theorem]{Claim}

\begin{document}

%\begin{flushright}
%\preprint{}
%\end{flushright}

\title{Explicit Bounds on the Spectrum of $6d$ $\mathcal N=(1,0)$ Supergravity}



\author{Caucher Birkar\vspace{2pt}} 
\email[]{birkar@tsinghua.edu.cn}
\affiliation{Yau Mathematical Sciences Center, Jingzhai, Tsinghua University, 
Hai Dian District, Bejing, China 100084}

\author{Seung-Joo Lee\vspace{2pt}}
\email[]{seungjoolee@yonsei.ac.kr}
\affiliation{Department of Physics, Yonsei University, Seoul 03722, Republic of Korea}
%\thanks{}

%\date{\today}

\begin{abstract}\vskip 3mm\noindent
We propose a novel strategy to derive explicit and uniform upper bounds on the particle spectrum of six-dimensional gravitational theories with minimal supersymmetry, focusing initially on the tensor sector. The strategy is motivated by considerations of F-theory compactifications on elliptic Calabi-Yau 3-folds. However, it admits a clear bottom-up interpretation and is thus applicable to general supergravity theories modulo certain physical conjectures. At the heart of the strategy are two key structures, most natural in birational geometry: one concerns the singularity of the natural pairs on the base manifolds, and the other, the fibration generically exhibited by the bases. Put physically, the former structure keeps the effective theories from decompactifying and the latter ensures the (generic) presence of a heterotic string. We sketch our bounding strategy and present, for an illustration, the explicit bounds thereby derived on the tensor spectrum, with the technical details relegated to a companion paper~\cite{ag}. 
%The geometric strategy developed in this note naturally generalizes to an analogous physical strategy, leading to universal spectrum bounds independent of specific top-down constructions, once certain physical assumptions are made. 

 \end{abstract}

%\pacs{}
%\keywords{}

\maketitle

%\section{}
%\subsection{}
%\subsubsection{}

%\tableofcontents

%%%%%%%%%%
\section{Introduction}\label{sec:intro}

The quantum nature of gravity has proven to impose general consistency constraints on the structure of effective theories. The precise origin of such constraints is oftentimes difficult to clarify via the field theoretic description of the effective physics alone. It is therefore of utmost interest to delineate the {\it landscape}, the set of effective theories with a consistent embedding into quantum gravity, whose complement, the {\it swampland}, is an infinite set of inconsistent theories~\cite{Vafa:2005ui}. 

As it turns out, many of the apparently-consistent effective theories end up being placed in the swampland, and hence, the notion of finiteness is naturally brought to the scene of the landscape. To be precise, the finiteness notion here concerns the discrete data of the effective theories, with the prototypical example being the particle spectrum, which counts light degrees of freedom.   

The general constraints commonly imposed on every consistent theory in the landscape are often proposed as a conjecture based on bottom-up intuitions.
 %Of the long list of conjectures that have been studied in the literature (see~\cite{} for a review), one of the earliest such conjectures is the Weak Gravity Conjecture~\cite{}, quantifying the weakness of gravity in various versions~\cite{}, and another is the Distance Conjecture~\cite{}, characterizing a common asymptotic feature of the theories at infinite distance, refined later on by the Emergent String Conjecture~\cite{}. 
For a complementary approach, it is thus important to analyze the individual conjectures from top-down perspectives, notably via string theory. In particular, one can pursue microscopic understanding of the conjectured constraints by revealing appropriate universal features of the internal compactification geometry, which strings are allowed to probe. In this context, enumerative geometry of Calabi-Yau varieties has served as a versatile toolkit for quantitative analyses while the role of birational geometry has not been as extensively studied by now. 

The aim of this note is to propose a novel strategy in birational geometry, by which one can derive explicit and uniform upper bounds on the supermultiplet spectrum of six-dimensional (6d) gravitational theories with $\mathcal N=(1,0)$ supersymmetry. The strategy is initially developed for the string vacua obtained via geometric compactifications of F-theory on an elliptic Calabi-Yau 3-fold. However, it is naturally interpreted purely in physical terms, indicating that the resulting bounds also constrain general effective theories modulo certain physical conjectures, without relying a priori on the geometric or the string-theoretic origin of the theories.  

While the finiteness of elliptic Calabi-Yau 3-folds (or equivalently, that of the geometric vacua of F-theory in six dimensions) was established decades ago~\cite{Gross},\footnote{To be precise, what was proven in~\cite{Gross} is that elliptic Calabi-Yau 3-folds form a bounded family modulo flops. The boundedness of topological types has been addressed more recently in~\cite{Filipazzi}; {see also~\cite{Grassi:2023aks} for some of the F-theoretic implications.}} no explicit bounds on their topological types and in particular on their Hodge numbers have so far been calculated. %by exploiting the singularity structure of natural pairs in the compact internal manifold, 
On the other hand, the strategy proposed in this note allows one to bound by concrete numbers some of the topological quantities as well as their physical counterparts, and manifests the origin of the bounds obtained. 
For an illustration, we will apply the strategy to bound the Picard number of the base 2-folds, thereby uniformly bounding the number $T$ of the 6d tensor multiplets. Such an explicit bound on the tensor sector will then serve as a stepping stone to the associated bounds on the other sectors of the theory.
%Interestingly, it turns out that we can naturally interpret the proposed geometric strategies in physical terms. This indicates that the resulting bounds may as well be derived without relying on the geometric structure of string vacua, modulo certain physical assumptions or conjectures. 

This note is organized as follows. In section~\ref{sec:review}, upon reviewing the basics of 6d $\mathcal N=(1,0)$ supergravity, we set the geometric arena up for the analysis of 6d F-theory and introduce the notion of pairs as well as their singularity types. %Crucially, for a given vacuum of 6d F-theory, the internal manifold and the 7-brane loci therein are put together to form a singular pair of the $\epsilon$-log canonical type with $\epsilon=1/6$. 
%, which results in explicit and uniform bounds on the spectrum of the associated 6d supergravity theories. 
Next, in section~\ref{sec:strategy}, we overview our bounding strategy and argue for explicit bounds on $T$, notably, 
%Furthermore, we also explain how an explicit tensor bound propagates to an associated vector bound. 
\beq\label{567}
T \leq {567} \,.
\eeq 
Here, the bound~\eqref{567} is by no means optimal, given that its derivation leaves much room for improvement~\cite{ag}, and should rather be viewed as a proof of principle. Finally, in section~\ref{sec:discussion}, we close with further discussion of physics, as well as comments on applications and generalizatios. \\[-.1in]

%This note serves as a brief summary of the proposed strategy, illustrating how it results in explicit spectrum bounds. The technical details will appear in the forthcoming paper~\cite{}, where improved bounds will also be presented.  

\noindent{\it Authors' note:} {\it having announced part of our results~\cite{FirstTalk} at the stage of completing the work, we learned of the highly relevant paper~\cite{Kim:2024hxe}, where the same physics problem was tackled.}

%%%%%%%%%%
\section{Background and Review}\label{sec:review}
The field contents of a 6d $\mathcal N = (1,0)$ supergravity theory are organized into super-multiplets of three types beyond the unique gravity multiplet. These are vector, tensor and hyper multiplets, whose numbers we denote respectively as $V$, $T$ and $H$. 

The (charged) hyper multiplets are the matter fields transforming under the gauge algebra $\frak g$ carried by the vector multiplets; for later purposes, we decompose $\frak g$ into the abelian and the non-abelian parts,  
\beq\label{g-split}
\mathfrak g=(\oplus_{\kappa} {\frak{u}(1)}_{\kappa}) \oplus (\oplus_\iota {\mathfrak g}_\iota)  \,,
\eeq 
where $\frak{g}_\iota$ are simple Lie algebras.
The gravitational and the gauge anomalies of the theory are characterized by the $\frak{so}(1,T)$ vectors known as the anomaly coefficients, 
\beq\label{abkbi}
a, b_{\kappa}, b_{\iota}  \in \mathbb R^{1,T} \,. 
\eeq
Here, the $T$ directions of $\mathbb R^{1,T}$ with a negative norm are associated to as many tensor multiplets, each containing one anti-chiral tensor along with a real scalar. 
Then, for the anomaly cancellation, various inner products of~\eqref{abkbi} are subject to certain consistency conditions, which, however, do not bound the spectrum by themselves~\cite{Kumar:2010ru, Park:2011wv, Taylor:2018khc}, rendering the set of anomaly-free theories infinite. 

Let us now consider the 6d supergravity theories with a geometric realization in F-theory, each described by a genus-one fibered Calabi-Yau 3-fold $\hat Y$,
\beq\label{hatpi}
\hat \pi: \hat Y \to Z \,, 
\eeq
whose base 2-fold $Z$ serves as the internal manifold. Such a fibration has an associated Weierstrass description,\footnote{If $\hat\pi$ does not have a section, we consider the Jacobian variety.}
\beq\label{pi}
\pi: Y \to Z \,,
\eeq
which we focus on in this note. Specifically, the Weierstrass model in scrutiny is of the form,
\beq\label{Weierstrass}
y^2=x^3+ F x + G \,,
\eeq
where $(F,G)$ is a pair of sections, 
\beq
F \in H^0(Z,  \mathcal O_Z(-4 K_Z)), ~~~ G \in H^0(Z, \mathcal O_Z(-6 K_{Z})) \,,
\eeq
with $K_{Z}$ denoting the canonical divisor of $Z$. Then, the singular fibers of $\pi$ are supported over the zero loci of the discriminant section, 
\beq\label{Delta}
\Delta:=4F^3+27 G^2 \in H^0(Z, \mathcal O_Z(-12K_{Z}))\,,
\eeq
whose irreducible components $\Delta_\iota$ wrapped by 7-branes carry the algebras $\mathfrak g_{\iota}$, as encoded in the vanishing orders, 
\beq
{\rm ord}_{\Delta_\iota} (F, G, \Delta) \,, 
\eeq
via the Kodaira-N\'eron classification. 

On the other hand, in algebraic geometry, one can naturally associate two divisors with the fibration $\pi$: one is the {\it discriminant divisor} $\Lambda$ characterizing the singularities, whether from the base, the fiber, or the total space, and the other is the {\it moduli divisor} $M$ defined by its divisor class $[M] = -[K_{Z}] - [\Lambda]$ with a certain positivity property. With the base $Z$ smooth, the combined divisor
\beq\label{B}
B:=\Lambda + M  \,,
\eeq
carries the same information\footnote{To be precise, the discriminant divisor $\Lambda$ agrees with $\frac{1}{12}\Delta$ over the irreducible components $\Delta_\iota$ supporting singular fibers {\it not} of the Kodaira type $I_1$, except that the components supporting $I_{n>1}$ and $I_{n>0}^*$ fibers, if any, lead to discrepancy; note that those exceptional loci only arise via a further fine tuning of the Weierstrass model. On the other hand, the moduli divisor $M$ is numerically equivalent to $\frac{1}{12} R$ for some effective integral divisor $R$, which, modulo the caveat stated, has the same divisor class as the $I_1$ locus.\label{f2}}  %{\red\raisebox{1ex}{\underline{\smash{\raisebox{-1ex}{, which can be chosen to represent the loci supporting $I_1$ fibers.}}}}}\label{f2}} 
as the divisor $\Delta$ in~\eqref{Delta}. 

Specifically, the discriminant divisor is determined as a linear combination of the irreducible divisors, 
\beq\label{Lambda}
\Lambda = \sum (1-t_C) C \,,
\eeq
where $t_C$ are the lc-threshold of $\pi^*C$ defined as 
\beq
t_C := {\rm sup}\{s\,|\,(Y, s\,\pi^*C)~{\text{is lc over general $z\in C$} } \}\,.
\eeq 
Here, {lc} refers to the singularity type known as {\it log canonical}, which we now turn to, restricting for simplicity to the current geometric setup. %; see e.g.~\cite{} for a general account. 

We start by introducing the notion of a {\it pair}, $(Z, B)$, which combines the smooth 2-fold $Z$ with an effective $\mathbb Q$-divisor $B$ such that $K_Z + B$ is $\mathbb Q$-Cartier; to clearly define singularities, we assume that $B$ has coefficients in $[0,1]$. For a log resolution $\phi: W \to Z$ of $(Z,B)$, we write 
\beq
K_W + B_W = \phi^*(K_Z+B)\,,
\eeq
where $B_W$ is uniquely determined, with the canonical divisor $K_W$ of $W$ chosen such that $\phi_*K_W = K_Z$. Then, we say that the pair $(Z,B)$ is {\it log canonical} (lc) if each coefficient of $B_W$ is not bigger than 1, and $\epsilon$-{\it log canonical} ($\epsilon$-lc) if not bigger than $1-\epsilon$ where $\epsilon \geq 0$.  
%%%%%%%%%%
\section{Overview of the Strategy}\label{sec:strategy}

%Note first that the numbers of tensor and vector multiplets, $T$ and $V$, amount respectively to the following topological invariants of the fibration~\eqref{pi} and its resolution~\eqref{hatpi}: 
%\bea\label{tensor-spec}
%T&=&h^{1,1}(Z)-1\,, \\ \label{vector-spec}
%V&=&h^{1,1}(\hat Y)-h^{1,1}(Y)\,. 
%\eea
%Here we will focus mostly on finding an explicit bound on $T$, or equivalently, on the Picard number $h^{1,1}(Z)$ of the base 2-fold. The bound on $T$ turns out to result in an associated bound on $V$, which we will briefly comment on in section.
In this section we sketch the birational strategy for deriving an upper bound on the tensor count $T$, which is related to the Picard number $h^{1,1}(Z)$ as 
\beq\label{tensor-spec}
T= h^{1,1}(Z)-1\,.  
\eeq
Mathematically, the goal is thus to bound $h^{1,1}(Z)$ from above. 
%As it turns out, the bound on $T$ propagates to associated bounds on the other sectors of the theory, which we will comment on in section~\ref{discuss-other-sectors}.
%Let us first note that the number of tensor multiplets is related to the Picard number of the base 2-fold, 
For readability, here we present the key ideas of the strategy, as well as some of the major claims, and proceed to an illustrative calculation of explicit bounds. Rigorous proofs, along with technical details, are left to the forthcoming paper~\cite{ag}. 
 
%%%%%
\subsection{{Key to the Boundedness of $T$}} \label{key-structures}
Two geometric structures lie at the heart of the bounding strategy, each bearing a clear physical interpretation. Let us elaborate on them in turn. 

%%%
\subsubsection{$\frac16$-lc singularity: general structure of $(Z,B)$}
The first key structure is found from the natural pair, 
\beq
(Z, B:=\Lambda+M)\,, 
\eeq
where $Z$ is the base 2-fold and the divisor $\Lambda$, along with $M$, encodes the fibral degenerations%, and hence, the 7-brane configurations
, as discussed in section~\ref{sec:review}. 
Then, the pair carries the singularity of a special kind: 
\beq\label{1/6lc}
\text{\it $(Z, B)$ is $\frac16$-lc\,.} 
\eeq
As observed in~\cite{ag}, the structure~\eqref{1/6lc} is inherent in the fibration~\eqref{pi} and has many interesting consequences, one of which is that every irreducible curve $C$ of $Z$ obeys  
\beq\label{-12}
C\cdot C \geq -12 \,.
\eeq

For a physical interpretation of~\eqref{1/6lc}, and in particular of~\eqref{-12}, we note that any curve $C$ with ${\rm coeff}_C(B) > \frac56$ or $C\cdot C < -12$, if present in $Z$, is necessarily subject to 
\beq
{\rm ord}_C (F, G, \Delta) \geq (4,6,12) \,, 
\eeq
indicating that the fibers over $C$ are not of a Kodaira type, often referred to as non-minimal fibers. Such non-minimal fibers over a codimension-one locus in the base obstruct Crepant resolutions of the Weierstrass model. Nevertheless, the model can be birationally modified in a systematic manner to be free of non-minimal fibers, as addressed firstly for the elliptic K3 surfaces~\cite{Lee:2021qkx}, and later on for the elliptic Calabi-Yau 3-folds~\cite{Alvarez-Garcia:2023gdd, ALW}; the fate of the codimension-one non-minimal fibers is thereby identified as the affinization of the gauge algebras, and in turn, as decompactifications~\cite{Lee:2021usk, Alvarez-Garcia:2023qqj}. 

In the current setup, the singularity structure~\eqref{1/6lc} can thus be interpreted physically as the obvious statement that we only consider supergravity theories in six dimensions, not in higher dimensions. 
%%%
\subsubsection{Rational fibration: generic structure of $Z$}
The other key structure is the $\mathbb P^1$ fibration exhibited by the generic base 2-folds. To be precise, the allowed base 2-folds $Z$ are classified into three types~\cite{Grassi}: the complex projective plane, the Enriques surface, and the Hirzebruch surfaces $\mathbb F_{a \leq 12}$ as well as their blowups. As the first two types have $h^{1,1}(Z)=1$ and $10$, which are faily small, however, we may assume the last type in bounding $h^{1,1}(Z)$. 
We thus consider an $r$-fold blowup of $\mathbb F_{a}$, 
\beq\label{ZBlr}
Z= {\rm Bl}^r(\mathbb F_a)\,, \quad r \geq 0 \,,
\eeq
with $h^{1,1}(Z) = 2+ r$ and seek after an upper bound on $r$.
In this case, $Z$ is a fibration,
\beq\label{P1fib}
p: Z \to \mathbb P^1_{\rm b} \,,  
\eeq 
whose generic fiber, 
\beq
f=p^{-1}(x)\,,\quad x \in \mathbb P^1_{\rm b}\,,
\eeq
is a smooth rational curve. 

While the singularity~\eqref{1/6lc} of $(Z,B)$ concerns the 7-brane physics, the fibration~\eqref{P1fib} of $Z$ is more relevant to the 3-brane physics. Notably, the D3-brane wrapped on $f$ leads to the effective {\it heterotic} string, which, at infinite distance in the tensor moduli space, serves as the weakly-coupled tensionless critical string~\cite{Lee:2018urn, Lee:2018spm}. While we are not particularly interested in the asymptotic physics here, the mere presence of a heterotic string, despite being tensionful at finite distance, turns out to play a pivotal role in bounding the spectrum. 

To fully exploit the fibration structure~\eqref{P1fib}, we make a dichotomy between the horizontal and the vertical parts of the divisor $B$,
\beq\label{hv}
B = B^{\rm h} + B^{\rm v} \,. 
\eeq
Here, the components of $B^{\rm h}$ intersect non-trivially with $f$, while those of $B^{\rm v}$ do not. Put physically, a 7-brane stack is horizontal if it is ``visible'' to the heterotic string, and vertical otherwise. 

%%%%%
\subsection{The Strategy and the Rationale} 
We begin by blowing $Z$ down to $Z_0 = \mathbb F_{a \leq 12}$, 
\beq
Z \xrightarrow{\varphi} Z_{0} \,, 
\eeq
after which we repeatedly blow $Z_0$ up as much as possible,  
\beq\label{chain}
Z_r \xrightarrow{\varphi_{r-1}} Z_{r-1} \xrightarrow{~~~~~} ~\cdots~ \xrightarrow{~~~~~} Z_1 \xrightarrow{~\varphi_0~} Z_0\,, 
\eeq 
where, by abuse of notation, the same symbol $r$ as in~\eqref{ZBlr} has been used. Here, each blowup $\varphi_i$ replaces a point $z_{i} \in Z_{i}$ by an exceptional divisor $E_{i+1} \subset Z_{i+1}$. We then show that $Z_r$ maps to the original base $Z$, which would imply in particular that 
\beq\label{h11Z-bound}
h^{1,1}(Z) \leq h^{1,1}(Z_r) = 2+ r\,.
\eeq
%Note that each map  appearing in the sequence~\eqref{chain} blows up a point $z_{i-1} \in Z_{i-1}$, resulting in an exceptional divisor $E_{i}\subseteq Z_{i}$. 

In order to find a uniform upper bound on $r$, we rely heavily on the $\frac16$-lc-ness of the pair $(Z, B)$. Importantly, this singularity structure is preserved under the birational transformations involved. In other words, each of the divisors $B_i$ of $Z_i$ defined by 
\bea
B_0&=& \varphi_*(B) \,,\\ 
K_{Z_i} + B_i &=& (\varphi_0 \circ \cdots \circ \varphi_{i-1})^*(K_{Z_0}+B_0) \,,
\eea
forms a $\frac16$-lc pair $(Z_i, B_i)$ {with $B_i$ effective}. Here, each $B_i$ also splits as
\beq
B_i = B_i^{\rm h} + B_i^{\rm v} \,,
\eeq
with respect to the fibration, 
\beq\label{pi-proj}
p_i: Z_i \to \mathbb P^1_{\rm b} \,, 
\eeq
induced from the $\mathbb P^1$-fibration $p_0$ of the Hirzebruch surface $Z_0$ as 
\beq
p_i = p_0 \circ \varphi_0 \circ \cdots \circ \varphi_{i-1} \,.
\eeq

Then, building essentially upon the two key structures~\eqref{1/6lc} and~\eqref{P1fib}, our strategy not only manifests the finiteness of the blowup chain~\eqref{chain}, but also results in an explicit upper bound on $r$, and hence, on $h^{1,1}(Z)$ via~\eqref{h11Z-bound}. In the following, for a glimpse of how such a bound could possibly arise, we present without proofs some of the major claims on the pairs $(Z_i, B_i)$, along with their consequences; a detailed account of the claims and their rigorous proofs, as well as a comprehensive description of the bounding strategy, can be found in~\cite{ag}.

%%%
\subsubsection*{Horizontal Multiplicities}

%Each blowup point $z_i$ lies either in one or two components of the fiber $p_i^{-1}(x)$ with $x:=p_i(z_i)$. The blowups of these two kinds will be called a {\it single} and a {\it double} blowup, respectively. 
%Furthermore, for a technical reason, 
As it turns out, the following multiplicities, 
\beq\label{order-param}
h_i:=\mu_{z_i} B_i^{\rm h}\,,   
\eeq 
which we refer to as the ``horizontal'' multiplicity of the respective blowup $\varphi_i$, play an important role in our bounding strategy. 
%\begin{itemize}
%\item Step 1 ($r_0 \leq i <r_1$):  ~  $m_1 \leq \mu_{z_i} B_i^{\rm h}$ %\quad for~ $0 \leq i < r_1$\,, 
%\item Step 2 ($r_1 \leq i < r_2)$: ~ $m_2 \leq \mu_{z_i} B_i^{\rm h}<m_1$ %\quad for~$r_1 \leq i < r_2$\,, 
%\item Step 3 ($r_2 \leq i < r_3)$: ~ $m_3\leq \mu_{z_i} B_i^{\rm h} <m_2$ %\quad for~$r_2 \leq i <r$\,.
%\end{itemize}
Let us assume without loss of generality that $\{h_i\}_{i=0}^{r-1}$ form a non-increasing sequence. Furthermore, we may suppose 
\beq
h_{r_1 -1} \geq \frac{1}{6} > h_{r_1} \quad \text{and} \quad h_{r_2-1} >  0 = h_{r_2} \,,   
\eeq
to subdivide the blowup chain~\eqref{chain} into three subchains, 
\bea\label{sc3}
Z_{r} &\xrightarrow{~\varphi_{r-1}}& Z_{r-1} ~\xrightarrow{~~~~~} ~\cdots~ \xrightarrow{~\varphi_{r_2}~} Z_{r_2} \\ \label{sc2}
&\xrightarrow{\varphi_{r_2-1}}& Z_{r_2-1} \xrightarrow{~~~~~}~ \cdots ~\xrightarrow{~\varphi_{r_1}~} Z_{r_1} \\ \label{sc1}
&\xrightarrow{\varphi_{r_1-1}}& Z_{r_1-1} \xrightarrow{~~~~~} ~\cdots~ \xrightarrow{\;~\varphi_0~\;} Z_0 \,.
\eea
%The notion of horizontal multiplicity thus serves as an order parameter for the subdivision of the blowups $\varphi_i$ in~\eqref{chain} into three subchains:  
%$h_i \geq \frac16$ for $i\in [0, r_1)$; $h_i \in (0, \frac16)$ for $i\in[r_1, r_2)$; $h_i = 0$ for $i\in[r_2, r)$. 
%\beq\label{subchains}
%  \begin{array}{ l l l l l l }
%Z_{r} &\to& Z_{r-1}& \to \cdots &\to& Z_{r_2} \\ 
%&\to& Z_{r_2-1}& \to \cdots &\to& Z_{r_1} \\ 
%&\to& Z_{r_1-1}& \to \cdots &\to& Z_0 \,.
%\end{array}
%\eeq
It follows that the above-defined notion of horizontal multiplicity provides intuitions behind how the total length $r$ of the chain is constrained, which we now turn to. 

%%%
\subsubsection*{Blowups with $h_i >0$}

To start with, the lengths of the subchains~\eqref{sc2} and~\eqref{sc1} are bounded via the following two claims. 

\begin{claim}\label{claim1}%For the blowups $\varphi_i$ in Step 1, 
The sum of the horizontal multiplicities obeys
\beq
\sum\limits_{i=0}^{r-1} h_i \leq 28 \,.
\eeq
\end{claim}

\begin{claim}\label{claim2}%For the blowups $\varphi_i$ in Step 1, 
Each blowup $\varphi_i$ with $h_i \in (0, \frac16)$ has in fact $h_i = \frac{1}{12}$. In other words, $h_i=\frac{1}{12}$ for ${}^\forall i \in [r_1, r_2)$. 
\end{claim}

\noindent Claim~\ref{claim2} assures that the multiplicities $h_i$ for $i<r_2$ cannot be arbitrarily small. This, once combined with Claim~\ref{claim1}, manifests how $r_2$, as well $r_1$, is bounded: 
\beq\label{r1r2-bound}
\frac{r_1}{6} + \frac{r_2-r_1}{12} \leq 28 \quad \Rightarrow \quad r_1 + r_2 \leq 336 \,.  
\eeq   

%%%
\subsubsection*{Blowups with $h_i =0$}

The next two claims below hint at the length bound for the remaining subchain~\eqref{sc3}.

\begin{claim}\label{claim3}
Each blowup point $z_i \in Z_i$ with $i \geq r_1$, i.e., with $h_i <\frac16$, lies at the intersection of a pair of curves, both vertical with respect to the fibration~\eqref{pi-proj}.
\end{claim}

\begin{claim}\label{claim4}
For any point $z \in Z_{r_2}$, the number, $\sigma_z$, of the blowups $\varphi_i$ with $h_i =0$ such that $(\varphi_{r_2} \circ \cdots \circ \varphi_i )(E_{i+1}) = z$ is subject to 
\beq\label{11}
\sigma_z \leq 11\,.
\eeq
\end{claim}
 
\noindent Note that $Z_{r_2}$ carries a total of $r_2$ points of vertical-curve intersection. Then, Claim~\ref{claim3} asserts that the subchain~\eqref{sc3} can only blow such points up, over each of which at best $11$ blowups may arise, according to Claim~\ref{claim4}. We thus have 
\beq\label{r-bound}
r- r_2 \leq 11\,r_2  \quad \Rightarrow \quad r \leq 12 \,r_2\,.
\eeq

%%%%%
\subsection{Calculation of Explicit Bounds}
Claims~\ref{claim1}--\ref{claim4} not only manifest the boundedness of $r$ but also lead to an explicit bound as 
\beq\label{conservative}
r \leq 12\, r_2 \leq 12\, (r_1 + r_2) \leq 4032 \,,  
\eeq
where~\eqref{r1r2-bound} and~\eqref{r-bound} have been used. 
However, the derived bound of $4032$ is rather conservative.  In particular, the first inequality in~\eqref{conservative} is only saturated if the subchain~\eqref{sc3} blows up the $r_2$ points of vertical-curve intersection in $Z_{r_2}$ maximally, each 11 times, and the second, if $r_1=0$, meaning that every blowup with $h_i >0$ in fact has $h_i=\frac{1}{12}$. 

Indeed, as shown in~\cite{ag}, we may improve~\eqref{conservative} to  
\beq\label{improved-r-bound}
r \leq 566 \,. 
\eeq  
by better constraining the number of ``small-$h_i$'' blowups $\varphi_i$ than how Claim~\ref{claim4} does. Specifically, to arrive at~\eqref{improved-r-bound} we need to not only sharpen Claim~\ref{claim4} on $\sigma_z$, concerning the blowups with $h_i=0$, but also constrain the number of blowups with $0<h_i <\frac16$ in a systematic manner. 
The both tasks start from looking closely into the structure of the {\it special fibers} of $p_r$ in $Z_r$, which have more than one components. 
In the rest of this section, for the illustrative purpose, we sketch how the former task of sharpening Claim~\ref{claim4} may be undertaken, relegating the detailed treatment of the both tasks to~\cite{ag}. 

To this end, let us associate with each special fiber $p_r^{-1}(x)$ over $x\in \mathbb P^1_{\rm b}$ the multiplicity, 
\beq
\mu_x^{\rm sf} := \mu_x \Lambda_{\mathbb P^1_{\rm b}} \,,  
\eeq
where 
$\Lambda_{\mathbb P^1_{\rm b}}$ is the discriminant divisor appearing in the canonical bundle formula, 
\beq
K_{Z_{r}} + B_{r} \equiv p^*_{r}(K_{\mathbb P^1_{\rm b}} + \Lambda_{\mathbb P^1_{\rm b}} + M_{\mathbb P^1_{\rm b}}) \,, 
\eeq
with the moduli part denoted by $M_{\mathbb P^1_{\rm b}}$.
%Having defined the multiplicity notion $\mu^{\rm sf}$ for special fibers, 
We proceed now to some extra constraints on the coefficients in $B_r$ of the vertical curves, which serve as a component of the special fibers with ``small multiplicities'': 
\begin{claim}\label{claim5}
For each component of the special fiber $p_r^{-1}(x)$ over $x \in \mathbb P^1_{\rm b}$, its coefficient $b$ in $B_r$ is subject to 
\beq\label{b-ineq}
b \leq 
\begin{cases}
    \frac{2}{3} &  \textrm{if } \mu_x^{\rm sf} < \frac34 \\[3pt]
    \frac{3}{4} &  \textrm{if } \mu_x^{\rm sf} < \frac56 
\end{cases}
\eeq  
\end{claim}
\noindent Notably, the coefficient bounds~\eqref{b-ineq} are more stringent than the general bound of $\frac56$ imposed by the $\frac16$-lc-ness of the pair $(Z_r, B_r)$. Then, they result in the following sharpened form of Claim~\ref{claim4}, again to do with the special fibers with small $\mu^{\rm sf}$: 
\begin{claim}\label{claim6}
For a point $z \in Z_{r_2}$, the number, $\sigma_z$, of blowups $\varphi_i$ with $h_i =0$ such that $(\varphi_{r_2} \circ \cdots \circ \varphi_i )(E_{i+1}) = z$ is bounded from above as
\beq\label{3-7}
  \sigma_z \leq \left\{ \arraycolsep=1.2pt\def\arraystretch{1.3}
  \begin{array}{ cl }
    3  & \quad \textrm{if } \mu_x^{\rm sf} < \frac34 \\
     5  &\quad \textrm{if } \mu_x^{\rm sf} < \frac56 
  \end{array}
\right.,
\eeq
where $x=p_{r_2}(z) \in \mathbb P^1_{\rm b}$.
\end{claim}

Claim~\ref{claim6} is particularly useful in improving the conservative bound~\eqref{conservative} on $r$, because $Z_{r}$ cannot carry too many special fibers with big $\mu^{\rm sf}$. In particular, one can show: 

\begin{claim}\label{claim7}%For the blowups $\varphi_i$ in Step 1, 
The fibration $p_r: Z_r \to \mathbb P^1_{\rm b}$ has at best two special fibers with $\mu_x^{\rm sf} \geq \frac34$, $x\in \mathbb P^1_{\rm b}$. 
\end{claim}
\noindent Interestingly, Claim~\ref{claim7} is a consequence of the degree bound, 
\beq\label{deg<=2}
{\rm deg}(\Lambda_{\mathbb P^1_{\rm b}}) \leq 2\,,  
\eeq
which reflects the compactness of the base $\mathbb P^1_{\rm b}$ of the fibration $p_r$.

Furthermore, one can also show that, of the total of $r_2$ points of vertical intersection in $Z_{r_2}$, only a limited number of points can lie on each individual fiber $p_{r_2}^{-1}(x)$ over a fixed $x\in \mathbb P^1_{\rm b}$. Such a fiber-wise constraint arises essentially from the following simple property: 
\begin{claim}\label{claim8}
For each $i$, the intersection of the fiber $f_i$ of $p_i:Z_i \to \mathbb P^1_{\rm b}$ and the horizontal divisor $B_i^{\rm h}$ is 2, i.e., 
\beq\label{fiBi}
f_i \cdot B_i^{\rm h} = 2 \,.
\eeq 
\end{claim}

\noindent Note that~\eqref{fiBi} may be considered as a compactness criterion along the fiber of $p_i$, in much the same way as~\eqref{deg<=2} is along the base. 

Then, it turns out that a set of constraints in the spirit of Claims~\ref{claim5}--\ref{claim8} eventually result in the improved bound~\eqref{improved-r-bound}, which extends also to the bounds on the Picard number and the tensor count as 
\beq\label{rPicT}
r \leq 566 \quad \Leftrightarrow 
\quad h^{1,1}(Z) \leq 568 \quad \Leftrightarrow\quad T \leq 567 \,. 
\eeq
While the naive bound~\eqref{conservative} has been significantly improved in~\eqref{improved-r-bound}, we emphasize that much room is still left for even further improvement, as will be clear from the detailed derivation of the latter in~\cite{ag}. 

Indeed it is our belief that one may derive   
\beq\label{rPicT-belief}
r \leq 192 \quad \Leftrightarrow 
\quad h^{1,1}(Z) \leq 194 \quad \Leftrightarrow\quad T \leq 193 \,,   
\eeq
which are saturated in concrete F-theory vacua~\cite{Aspinwall:1997ye, Candelas:1997eh}, but which have remained a conjecture~\cite{Morrison:2012js} for long. 
The strongest bounds~\eqref{rPicT-belief} can in fact be established physically via bottom-up ingredients, as recently pointed out in~\cite{Kim:2024hxe}, where various anomaly constraints among others play a central role in arguing for the abstract boundedness to start with. 

In this regard, the virtue of our strategy is that it manifests both the abstract and the explicit boundedness, essentially from a handful of principal structures, namely,~\eqref{1/6lc} and~\eqref{P1fib}, which, as discussed, are interpretable purely in physical terms. While there is no doubt from the physical perspective that anomalies must cancel, their geometric incarnation has not been fully addressed.\footnote{Non-trivial evidence is found e.g. in~\cite{Grassi:2011hq, Grassi:2018rva}, providing partial proofs from the geometric perspective in special cases.} It is thus worth emphasizing that the way our strategy leads to the bounds~\eqref{rPicT} still leaves much room for an optimization without directly relying on the anomaly considerations. 

%%%%%%%%%%
\section{Summary and Discussion}\label{sec:discussion}
In this short note, we have overviewed a novel strategy for bounding the number $T$ of tensor multiplets in compactifications of F-theory to six dimensions. Exploiting the strategy, we have also derived an explicit upper bound~\eqref{rPicT} on $T$, which uniformly applies to all 6d effective theories of F-theory. 

Geometrically, the defining data for a 6d F-theory vacuum consists of a complex surface $Z$ and an elliptic fibration $\pi: Y\to Z$, with a Calabi-Yau total space $Y$. Here, the singularities of $Y$ are encoded in a certain divisor $B$ of $Z$ with $K_Z + B \equiv 0$, as defined in~\eqref{B}, so that the pair $(Z, B)$ may naturally label the effective theory in scrutiny. For the purpose of bounding $h^{1,1}(Z)=T+1$, we have then focused on the two key structures: 
\begin{enumerate}
\item ${\rm Singularity~\eqref{1/6lc}:~}(Z,B)~\text{is $\frac16$-lc};$ %\\  \label{key2}
\item ${\rm Fibration~\eqref{P1fib}:~} Z~\text{is fibered with generic $\mathbb P^1$ fibers}.$
\end{enumerate}
%which have been singled out in this note for the purpose of bounding $h^{1,1}(Z)=T+1$. 
While the former structure is found in a general F-theory vacuum, the latter is exhibited only generically, i.e., when the base surface $Z$ is not $\mathbb P^2$ nor the Enriques surface; the two exceptional bases can, however, be safely ignored in view of our bounding task. 
We have then learnt that essentially these two simple structures severely constrain the geometry of F-theory vacua in a universal manner, leading to an explicit bound on $h^{1,1}(Z)$. To make full use of the both structures, we have considred a two-fold split of the divisor $B$, the key player for the $\frac16$-lc-ness, into the horizontal and the vertical parts, with respect to the fibration~\eqref{P1fib}.

Specifically, beginning by blowing $Z$ down to a Hirzebruch surface $Z_0=\mathbb F_a$, we have blown the latter up as much as possible, eventually arriving at $Z_r$, not allowed to be further blown up. Crucially, both of the key structures are preserved along the entire chain~\eqref{chain} of birational transformations involved, where at each stage, a pair $(Z_i, B_i)$, as well as the two-fold split of $B_i$, is naturally defined. The spirit of our bounding strategy is then to appropriately subdivide the blowup chain~\eqref{chain} into subchains, e.g., into~\eqref{sc3}--\eqref{sc1}, and to constrain each subchain. For the subdivision of this kind, the horizontal multiplicity, $h_i$, as defined in~\eqref{order-param}, naturally serves as an order parameter.

%%%%%
%\subsection{Physical Interpretation}\label{discuss-physics}
The proposed strategy is all the more interesting in that it is naturally interpreted in physical terms. In particular, each of the two key structures playing a pivotal role in the derivation of~\eqref{rPicT} results from a simple physical principle as follows:    %To this end, some of the physical constraints and/or conjectures may have to be imposed on top. 
\begin{enumerate}
\item Absence of affine gauge algebras $\Rightarrow (Z,B)$ is $\frac16$-lc; %\\  \label{key2}
\item Presence of a critical string $\Rightarrow$ $Z$ is a fibration. 
\end{enumerate}
%Let us elaborate on the two principles in turn. 

Firstly, the $\frac16$-lc structure is imposed by banning the affinization of gauge algebras. Already from the bottom-up perspective, any affine enhancement involves an infinite tower of massless states, indicating that the effective description of the theory breaks down.  Furthermore, a quantum gravity expectation is that those states are organized into a light Kaluza-Klein tower, triggering decompactification of the external spacetime~\cite{Alvarez-Garcia:2023qqj}; we may thus impose the absence of affinized algebras when constraining the theories in a fixed spacetime dimension.\footnote{For the earlier analyses of decompactifications via an affine enhancement, see e.g.~\cite{Lee:2021usk, Font:2020rsk, Collazuol:2022oey, Cachazo:2000ey}, where twice as many conserved supercharges are assumed.}

Next, the fibration structure follows from the presence of a (dual) critical string. A physical motivation for the latter is provided by the Emergent String Conjecture~\cite{Lee:2019wij}, which posits that a weakly-coupled string emerges in the asymptotic boundary of the moduli space, unless the spacetime decompactifies. We may thus impose that all the theories in scrutiny posses a critical string, which becomes tensionless at infinite distance to serve as a predicted weakly-coupled string. In the current F-theoretic setup, this implies that $Z$ is a fibration with the generic fiber either rational or elliptic, respectively, resulting in a heterotic or a type II string. However, the latter implies that no 7-branes are present and hence, that $Z$ is an Enrique surface. For completeness, we must also consider the case where no moduli except for the overall volume modulus are available; no fibration structures are required in this case, for which $\mathbb P^2$ is singled out as the only other exceptional base surface. 

The individual geometric claims~\ref{claim1}--~\ref{claim8} may be thought of as imposing global constraints on the conformal matter~\cite{DelZotto:2014hpa}, thereby limiting the superconformal field theories and the little string theories that are embeddable into a quantum gravity theory. In other words, many of those constraints would have not arisen had it not been for gravity, i.e., had the internal manifold $Z$ been non-compact. 
Notably, the bounding task as a geometric problem from the top-down perspective allows us to impose outright both the singularity and the fibration structures in particular. However, we may as well argue for the derived bounds also from the bottom-up perspective, given that each of the two key structures follows from an underlying physical principle as discussed. 

In this regard, it is fascinating to notice that the bottom-up principles behind are tied precisely with the two universal {\it infinite-distance} phenomena, as characterized by the Emergent String Conjecture; surprisingly, they turn out to severely constrain the {\it finite-distance} physics as well. The mere presence of a heterotic string leads to the useful dichotomy between the ``horizontal'' and the ``vertical'' gauge sectors, not necessarily relying on its top-down realization via geometry as a solitonic string. 
Furthermore, the absence of affine gauge algebras, on top of the anomaly cancellation, is an unquestioned prerequisite for the classification in~\cite{Heckman:2015bfa, Bhardwaj:2015oru} of superconformal field theories and little string theories; the outcome of such classification efforts then serves in~\cite{Kim:2024hxe} as one of the crucial ingredients for the strong bound~\eqref{rPicT-belief} on $T$, along with other spectral bounds. While the bound~\eqref{rPicT} derived in this note for the purpose of illustrating our bounding strategy is somewhat weaker, we emphasize again that there is much room left for an improvement without bringing in additional principles such as the anomaly cancellation. We leave it to future work to clarify the precise connection between the two approaches.  

We end with further comments on applications and generalizations of our results.  

%%%%%
\subsection{Bounds on the Other Sectors}\label{discuss-other-sectors}
An explicit and uniform bound on $T$ naturally serves as a starting point for bounding the other sectors of the theory, i.e., the vector and the hyper multiplet sectors, which we briefly discuss in turn. 

%%%
\subsubsection*{Vector Multiplet Sector}

%We first recall how the vector multiplet sector arises from the elliptic fibration $\pi$ in~\eqref{pi}. 
The geometric origin of the gauge algebra $\frak g$ in~\eqref{g-split} is rather different for the abelian and the non-abelian parts. 
The abelian vectors are in one-to-one correspondence with the non-zero sections of the elliptic fibration $\pi$ in~\eqref{pi}, and hence, are counted as
\beq\label{rk0}
R_0 := {\rm rk}({\rm MW}(\pi)) \,,
\eeq
where ${\rm MW}(\pi)$ denotes the Mordell-Weil group of rational sections. On the other hand, the non-abelian are tied with the Crepant resolution, $\hat Y \to Y$, of the singular Weierstrass model~\eqref{pi} to~\eqref{hatpi}.\footnote{While we assume, for simplicity of presentation, that $\hat Y$ is smooth, $\mathbb Q$-factorial terminal singularities may in principle remain, in which case~\eqref{neutral-hyper} receives corrections from Milnor numbers~\cite{Arras:2016evy}. Our bounding strategy still persists, however.} In particular, the resolution divisors are responsible for the Cartan subalgebra of each simple algebra $\frak g_\iota$, so that the combined non-abelian rank $\hat R := \sum_\iota {\rm rk}(\frak{g}_\iota)$ is given as  
\beq\label{hatrk}
\hat R = h^{1,1}(\hat Y) - h^{1,1}(Y) \,, 
\eeq
where $h^{1,1}(Y)$ represents the following sum,  
\beq
h^{1,1}(Y) = h^{1,1}(Z) + R_0 + 1 \,, 
\eeq
with $R_0$ defined in~\eqref{rk0}~\cite{TSW}. 

Let us first consider the full rank of $\frak{g}$, 
\beq\label{full-r}
R := {\rm rk} ({\frak g}) = R_0 + \hat R\,.  
\eeq
which receives separate contributions~\eqref{rk0} and~\eqref{hatrk}, respectively, from the abelian and the non-abelian vectors. 
With the former subject to a known bound~\cite{Lee:2019skh, Lee:2022swr}, 
\beq\label{r0-bound}
R_0 \leq 18 \,,
\eeq
here we sketch how the latter could be bounded.  
To this end we observe that the Cartier divisor, 
\beq
12B =: \bar\Delta = \sum \bar \delta_\iota \bar \Delta_\iota\,, 
\eeq
which is essentially\footnote{See footnote~\ref{f2} for the potential discrepancy.} the zero locus of the section $\Delta$ in~\eqref{Delta}, has each coefficient $\bar \delta_\iota$ bounded. Knowing how $h^{1,1}(Z)$ is bounded, we may then also bound $\sum \bar\delta_\iota$, thereby obtaining an explicit bound on $\hat R$, and hence, on $R$ via~\eqref{r0-bound}.

Furthermore, given the rank bound, we may also bound the number $V$ of vector multiplets, which amounts to the dimension of the gauge algebra $\mathfrak g$, splitting as 
\beq
V :=V_0 +  \hat V \,, 
\eeq 
where $V_0 = R_0$ and $\hat V = \sum_\iota {\rm dim}(\frak g_\iota)$ are the dimensions of the abelian and the non-abelian sectors, respectively. 
Here, the individual dimensions ${\rm dim}(\frak g_\iota)$ are bounded, given the associated rank bounds. An explicit bound thus arises on $\hat V$, and hence, on $V$; we leave the calculation of such an explicit bound for future work. 

%%%
\subsubsection*{Hyper Multiplet Sector}
Once explicit bounds are imposed on $T$ and $V$, what remains to be bounded is the number $H$ of hyper multiplets. We split $H$ as 
\beq
H = H_0 + (H-H_0)\,, 
\eeq
where $H_0$ counts the uncharged matter fields and connects to the deformations of $\hat Y$ via
\beq\label{neutral-hyper}
H_0 = h^{1,2}(\hat Y) + 1\,. 
\eeq
On the other hand, the remaining contribution $H-H_0$ from the charged fields is related to the fibral singularities over isolated points in the base $Z$. 

Note that $H_0$ in~\eqref{neutral-hyper} indicates that $H$ is not entirely birational in nature, unlike $T$ and $V$, as it involves counting of 3-cycles. In the meantime, physically, the gravitational anomaly constrains $H$ as 
\beq
H= 273 - 29T + V\,,  
\eeq
which, once combined with given bounds on $T$ and $V$, leads to an associated explicit bound on $H$ too. 
We leave an independent geometric exploration of the bound on $H$ for future work.

%%%%%
\subsection{Generalizations}\label{discuss-gen}

As for a most natural generalization, we may consider tackling the analogous problem of bounding $h^{1,1}(Z)$ in the case with ${\rm dim}(Z)=3$, where $Z$ serves as the base 3-fold of an elliptic Calabi-Yau variety, again with only Kodaira type fibers. This new task is much harder, given the richer geometric structure. 
{Building upon recent advances in birational geometry~\cite{g1, g2, g3}, however, one can attempt to generalize our bounding strategy, thereby establishing a similar, explicit boundedness notion for the 4d gravitational theories with $\mathcal N=1$ supersymmetry.
%Notably, the specific anomaly structure in 6d, based on which spectral bounds are physically derived in~\cite{Kim:2024hxe}, no longer holds in 4d. 
Notably, the $\frac16$-lc structure may still be imposed to prohibit any affine gauge algebras. Furthermore, the fibration structure with generic $\mathbb P^1$ fibers persists to a great extent; see~\cite{g4}. 
It is thus particularly exciting that the key ideas of our bounding strategy are still applicable to the base 3-folds. We leave this as an interesting future direction of investigation. }

Let us also note that the fibration $\pi$ in~\eqref{pi} does not play any significant role in bounding $h^{1,1}(Z)$; while the two key structures~\eqref{1/6lc} and~\eqref{P1fib} follow from $\pi$, in no later steps do we refer back to it.  
We may thus consider generalizing the bounding strategy to the case where~\eqref{1/6lc} is replaced by $\epsilon$-lc-ness with $\epsilon \neq \frac16$. Obviously, there should arise an explicit bound on $h^{1,1}(Z)$ for each chosen value of $\epsilon$. Despite that its physical interpretation would not be as clear, the task of revealing the $\epsilon$-dependence of the bound serves as an interesting geometrical problem per se.   \vspace{2mm}\\
%Moreover, it is natural to also consider exploiting the notion of generalized pairs~\cite{}.  

%%%%%%%%%%%%%%%%%%%%%%%%%%%%%%%%%%%%%%%

{\noindent \bf Acknowledgements}
%%%%%%%%%%
%\section*{ACKNOWLEDGEMENTS} 
S.-J.L. is grateful to Dongseon Hwang and Timo Weigand for useful discussions. The work of S.-J.L. is supported by the Yonsei University Research Fund 2025-22-0134. C.B. is supported by a grant from Tsinghua University and a grant from National Program of Overseas High Level Talent.

%%%%%%%%%%
%\section*{\red Comments/Questions}
%{\red
%\begin{itemize}
%\end{itemize}
%}

\end{document}